\documentclass[aps,prl,twocolumn,superscriptaddress, showpacs]{revtex4-1}
\usepackage{amssymb,graphicx,xspace,color}
\usepackage[pdfauthor={Frank Steckel}, bookmarks=true, pdfborder={0 0 0}, colorlinks=true,citecolor=blue,linkcolor=black,urlcolor=blue]{hyperref}
\newcommand{\tn}{$T_N$\xspace}

\newcommand{\lac}{La$_2$CuO$_4$\xspace}
\newcommand{\sio}{$\rm Sr_2IrO_4$\xspace}

\begin{document}
\title{Pseudospin heat conductivity in the $J_\mathrm{eff}=1/2$ antiferromagnet Sr$_2$IrO$_4$}


\author{Frank Steckel}
\email[]{f.steckel@ifw-dresden.de}
\affiliation{Leibniz-Institute for Solid State and Materials Research, IFW-Dresden, 01069 Dresden, Germany}
\author{Akiyo Matsumoto}
\affiliation{Department of Physics and Department of Advanced Materials, University of Tokyo, Hongo 113-0033, Japan}
\author{Tomohiro Takayama}
\affiliation{Max-Planck-Institute for Solid State Research, 70569 Stuttgart, Germany}
\author{Hidenori Tagaki}
\affiliation{Department of Physics and Department of Advanced Materials, University of Tokyo, Hongo 113-0033, Japan}
\affiliation{Max-Planck-Institute for Solid State Research, 70569 Stuttgart, Germany}
\author{Bernd B\"uchner}
\author{Christian Hess}
\email[]{c.hess@ifw-dresden.de}
\affiliation{Leibniz-Institute for Solid State and Materials Research, IFW-Dresden, 01069 Dresden, Germany}
\affiliation{Center for Transport and Devices, TU Dresden, 01069 Dresden, Germany}
\date{\today}
\begin{abstract}
We report the in-plane and out-of-plane heat conductivity of the antiferromagnetic spin-orbit induced Mott insulator \sio with $J_\mathrm{eff}=1/2$. Our data reveal clear-cut evidence for magnetic heat transport within the IrO$_2$ planes which provides the unique possibility to analyze the thermal occupation and scattering of $J_\mathrm{eff}=1/2$ pseudospin excitations. The analysis of the magnetic heat conductivity yields a low-temperature ($T\lesssim75$~K)
magnetic mean free path $l_\mathrm{mag}\approx  32$~nm, consistent with boundary scattering. Upon heating towards room temperature, the mean free path strongly decreases  by one order of magnitude due to thermally activated scattering of the pseudospin excitations. The latter reveals that the coupling of these excitations to the lattice is  radically different from that of $S=1/2$-excitations in cuprate analogs.
\end{abstract}

%\keywords{}
\maketitle
The physics of iridium oxide materials has recently moved into the focus as these materials realize a plethora of novel quantum magnetic phases based on, e.g., the square, honey-comb, and hyperkagome lattice types \cite{Kim2009,Kim2008,Ye2013,Singh2010,Okamoto2007} with $J_\mathrm{eff}=1/2$ pseudospins. This includes the sought-after possible realization of a quantum spin-liquid with peculiar elementary excitations \cite{Qi2009,Lawler2008}. The magnetic heat conductivity is considered an important tool to probe quantum spin and topological excitations \cite{Qi2009}, as it is sensitive to both the thermal occupation and the scattering of such quasiparticles. In the past years, this sensitivity has been exploited extensively for probing the elementary spin excitations in many different $S=1/2$ low-dimensional quantum magnets \cite{Yamashita2010,Yamashita2009,Sologubenko2001,Hess2007,Hlubek2010,Sologubenko2000,Hess2001,Hess2003,Berggold2006,Hess2007b}.
For materials with a strong spin-orbit coupling (SOC) and $J_\mathrm{eff}=1/2$ pseudospins, such as the iridates, it remains however unclear whether heat transport can be used to probe the pseudospin excitations at all, because  the strong SOC is likely to cause strong scattering due to phonons. Accordingly, successful experiments on iridate materials are lacking, apart from pioneering attempts \cite{Singh2013}.

One of the up to the present best studied iridate materials is the compound \sio which is a spin-orbit induced Mott insulator \cite{Kim2008} with localized electrons on the Ir$^{4+}$ ions in a $J_\mathrm{eff}=1/2$ state. The material possesses a very similar structure as  La$_2$CuO$_4$, i.e. a square lattice of Ir$^{4+}$ ions is formed by corner-sharing IrO$_2$ plaquettes, where adjacent IrO$_2$-planes are separated by SrO layers \cite{Huang1994}. A strong antiferromagnetic exchange of the order $J\sim0.06$~eV couples the $J_\mathrm{eff}=1/2$ pseudospins giving rise to two-dimensional (2D) magnetic excitations as is revealed by resonant inelastic x-ray scattering (RIXS) \cite{Kim2012}. \sio orders long-range antiferromagnetically at $T_N\approx224$~K, where a  weak ferromagnetic moment occurs due to canting of the IrO$_6$ octahedra \cite{Cao1998,Jackeli2009,Ye2013}.

In this Letter, we report the in-plane and out-of-plane heat conductivity of \sio. We observe a highly unusual in-plane heat conductivity with anomalous temperature dependence at $T\lesssim T_N$ which is absent for the out-of-plane direction. This is incompatible with phonon heat transport and evidences magnetic heat transport within the IrO$_2$ planes. 
Thus, our data reveal the first example for magnetic heat transport in a $J_\mathrm{eff}=1/2$ compound.
We analyze the low temperature magnetic heat conductivity $\kappa_\mathrm{mag}$ in terms of a Boltzmann-type approach and extract the magnetic mean free path $l_\mathrm{mag}$. At low temperature, $l_\mathrm{mag}$ is limited by boundary scattering. Upon heating to room temperature, $l_\mathrm{mag}$ decreases by an order of magnitude, indicating temperature-activated scattering. This scattering process is the dominating one at elevated temperature and can be assigned to strong magneto-elastic coupling.
% An expectable influence of the spin-spin correlation length is rendered of minor importance as in the considered temperature range the latter is nearly two orders of magnitude larger than $l_\mathrm{mag}$.

The growth and characterization of single crystals of \sio have been described in Ref.~\onlinecite{Kim2009}. The crystal dimension in our experiment was $0.51 \times 0.87 \times 0.17$~mm$^3$ where the shortest edge is the crystallographic $c$-direction. The thermal conductivity $\kappa$ has been measured with a home-made device in a four-point configuration using a chip resistor as heater and a thermocouple to measure the temperature gradient parallel to the $ab$-planes \cite{Hess2003a}. Due to the limited size of the thin plate-like single crystal it was impossible to perform a four-point measurement along the $c$-direction. Nevertheless, a two-point measurement was possible. For our setup we carefully investigated the differences between the two- and four-point configuration. For the two-point configuration we found, that above $\sim 150$~K the temperature dependence of the heat conductivity is reproduced correctly with the caveat of an uncertain absolute value. Therefore, we consider data from the two-point measurement only to search for anomalous temperature dependence in the vicinity of \tn.

\begin{figure}
\includegraphics[clip,width=1\columnwidth]{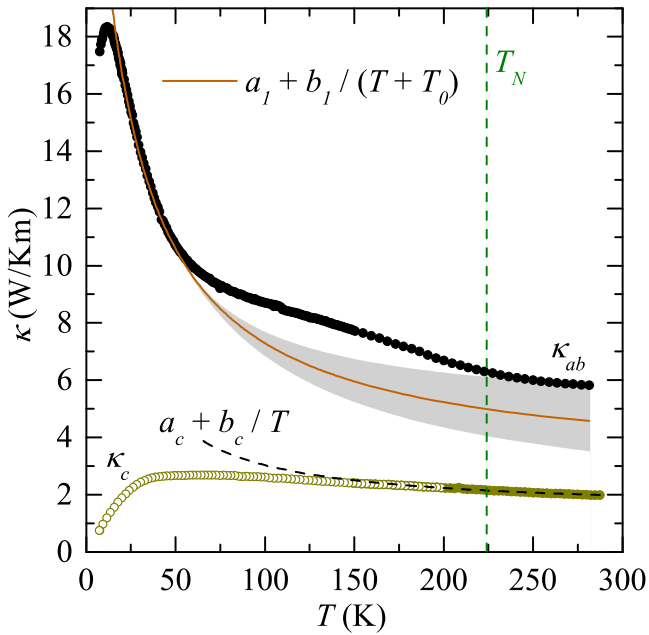}
\caption{Heat conductivity $\kappa$ of \sio measured in the $ab$ plane ($\kappa_{ab}$) and along the $c$ direction ($\kappa_{c}$). Purely phononic fits to the heat conductivity are shown as lines. The fit to the $ab$ direction is given by the values $a_1 =2.8$~W/Km, $b_1 =530$~W/m and $T_0 =18$~K. Similarly, the phononic fit to the $c$ direction with $a_c =1.43$~W/Km, $b_c =160$~W/m. For the $ab$ direction extreme fit parameters were used to find limits of the phononic temperature behavior marked as the shaded area. The upper bound is found by the values $a_0 =4.811$~W/Km, $b_0 =284.4$~W/m and $T_0=0$, the lower bound by $a_2 =1.2$~W/Km, $b_2 =700$~W/m and $T_0=25K$.} 
\label{fig:figure1}
\end{figure}

Fig.~\ref{fig:figure1} shows the measured heat conductivity of \sio along the $ab$ and $c$ directions, $\kappa_{ab}$ and $\kappa_{c}$, respectively. From resistivity measurements \cite{Korneta2010, Kini2006} and applying the Wiedemann-Franz-law it follows that \sio has a low contribution of electrons to the heat conductivity 
% of maximum $\kappa_{el}\sim7\cdot10^{-3}$~W/Km 
which is three orders of magnitude lower than the measured $\kappa$. Therefore, we neglect the contribution of the electrons and consider the total heat conductivity in $ab$ direction as the sum of a conventional phononic and a potential magnetic contribution, whereas $\kappa_c$ is purely phononic.

The in-plane heat conductivity $\kappa_{ab}$ exhibits a peak at low temperature ($\sim 12$~K), which is followed by a steep decrease that slightly levels off at around 75~K. At further increased temperature a broad step around \tn is observed and the curve almost saturates close to room temperature. This temperature dependence is incompatible with canonical phononic heat conduction.
In a simple approach the heat conductivity is proportional to the specific heat $c_V$, the velocity $v$, and the mean free path $l$ of the heat carriers: 
\begin{equation}
\kappa\sim c_Vvl.
\label{eq:cvl}
\end{equation}
In the case of phonons as heat carriers, the velocity and the mean free path are approximately constant at low temperature and $\kappa$ follows the temperature dependence of the specific heat. At higher temperatures, umklapp scattering becomes important which reduces the mean free path and thus leads to the observed low-temperature peak. This process depends on the number of excited phonons and leads to $l\propto 1/T$. Thus, at high temperatures where $c_V$ approaches the Dulong-Petit constant, the phononic heat conductivity is approximated by $\kappa_\mathrm{phon}\propto 1/T$ \cite{Berman1976}.

The leveling off at $\sim 75$~K and the broad step-like feature near \tn are clearly inconsistent with this expected temperature dependence. Two completely different scenarios are conceivable for explaining this unexpected behavior. It is possible that enhanced scattering of phonons occurs due to critical magnetic fluctuations near \tn. In fact, a dip structure near \tn is often found in antiferromagnetic materials like in MnO \cite{Slack1958} or in CoF$_2$ \cite{Slack1961}. However, such critical fluctuations are unlikely to affect $\kappa_\mathrm{phon}$ near 75~K, i.e. far away from \tn. Moreover, the phonon scattering due to magnetic fluctuations typically affects the heat conductivity isotropically even in layered systems \cite{Hofmann2001}. Therefore, we carefully inspected the heat conductivity parallel to the $c$-axis in the vicinity of \tn. At temperatures higher than $\sim$200~K, the $\kappa_c$ curve is absolutely featureless and fully described by the aforementioned $1/T$-law as is indicated in the figure. Thus, we can clearly rule out a phononic origin of the anomalous behavior in $\kappa_{ab}$. On the other hand, the anomalous behavior can also arise from a 2D magnetic heat conductivity within the IrO$_2$-layers which adds to $\kappa_\mathrm{phon}$, i.e., $\kappa_{ab}$ results from the sum of phononic and magnetic contributions while $\kappa_{c}$ is purely phononic. Indeed, such magnetic heat transport is frequently observed in low-dimensional $S=1/2$ quantum magnets \cite{Yamashita2010,Yamashita2009,Sologubenko2001,Hess2007,Hlubek2010,Sologubenko2000,Hess2001,Hess2003,Berggold2006,Hess2007b}. Thus, we conclude that in \sio  a magnetic contribution to the heat conductivity is present in $\kappa_{ab}$.

Having established this main experimental finding, namely the first observation of magnetic heat conductivity in a $J_\mathrm{eff}=1/2$ system, we move on to its quantitative analysis by extracting the magnetic mean free path. A phenomenological $T^{-1}$ approach is used to model the phononic heat conductivity at high temperatures (cf. the solid line in Fig.~\ref{fig:figure1}). To estimate the uncertainty of the phononic background in $\kappa_{ab}$, we performed extreme phononic fits for determining lower and upper bounds as is indicated by the shaded area \footnote{Note that the lower bound found from a $1/(T-T_0)$ fit resembles the high temperature behavior found by a fit to the Callaway model \cite{Callaway1959}.}.

\begin{figure}
\includegraphics[clip,width=1\columnwidth]{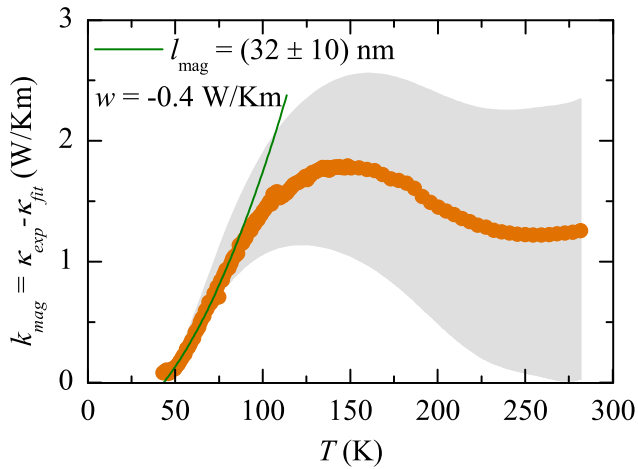}
\caption{Magnetic heat conductivity $\kappa_\mathrm{mag}$ of \sio resulting from the difference between $\kappa$ and the phononic fit (dots). The line is the low-temperature fit with constant magnetic mean free path $l_\mathrm{mag}$ and a constant shift $w$ from Eq.~(\ref{eq:1}). The uncertainty from the phononic fit is spanned by the shaded area.}
\label{fig:figure2}
\end{figure}

In the temperature range between \tn and $\sim50$~K the measured $\kappa$ exceeds the expected phononic heat conductivity remarkably, corroborating our conclusion of a significant $\kappa_\mathrm{mag}$. We subtracted the phononic fit from the measured $\kappa$ and obtain $\kappa_\mathrm{mag}$ as shown in Fig.~\ref{fig:figure2}. $\kappa_\mathrm{mag}$ increases with increasing temperature up to $\sim 150$~K with a maximum value of about $1.8$~W/Km, and decreases for higher temperatures. If one considers the coarse generic behavior of heat conductivity given by Eq.~(\ref{eq:cvl}), the low-temperature increase arises from the thermal occupation of pseudospin excitations (termed magnons hereafter). The peak at 150~K and the high-temperature decrease of $\kappa_\mathrm{mag}$ cannot be related to the maximum of the specific heat of 2D spin excitations, as the latter is expected around $0.6\, J/k_B\sim420$~K \cite{Sengupta2003} (with $J$ the exchange constant in a 2D $S=1/2$ Heisenberg model), i.e. at much higher temperatures than considered here. Instead, the decrease must be primarily related to a temperature dependence of the mean free path since the average magnon velocity is unlikely to change strongly at the rather low temperatures ($T\ll J/k_B$, with $J=0.06$~eV \cite{Cetin2012,Hirata2013,Kim2012}) considered here.

For investigating our result for $\kappa_\mathrm{mag}$ further, we follow an approach that has previously been used to analyze $\kappa_\mathrm{mag}$ of the $S=1/2$ analog La$_2$CuO$_4$  \cite{Hess2003}, i.e., we extend the simple kinetic description of Eq.~(\ref{eq:cvl}) by accounting for possible momentum dependencies in two dimensions \cite{Hess2003,Hess2007b}, i.e.
$
 \kappa\propto\int c_{\bf k}v_{\bf k}l_{\bf k}d{\bf k}, \label{kinetic}
$
with $c_{\bf k}=\frac{d}{dT}\epsilon_{\bf k}n_{\bf k}$ the specific heat ($\epsilon_{\bf k}$ and $n_{\bf k}$ are the energy and the Bose occupation function of the mode $\bf k$), $v_{\bf k}$ the velocity and $l_{\bf k}$ the mean free path of a magnon with wave vector ${\bf k}$. 

In \sio, the dispersion $\epsilon_{\bf k}$ is a steeply increasing function with band maxima at $\epsilon>100~{\rm meV}= k_B\cdot 1160~{\rm K}$ \cite{Kim2012}. Thus, at the low temperatures considered here, primarily modes with small momenta are relevant for the heat transport. For simplicity, we therefore assume a temperature independent mean free path $l_\mathrm{mag}\equiv l_{\bf k}$. We approximate the dispersion $\epsilon_{\bf k}=\epsilon_k=\sqrt{\Delta^2+(\hbar v_0 k)^2}$ \cite{Hess2003}, where $\Delta=0.83$~meV is the experimental anisotropy gap revealed by ESR \cite{Bahr2014} and $v_0=(6.25 \pm 0.4)\cdot 10^4$~m/s the small-$k$ magnon velocity extracted from the RIXS single magnon dispersion \cite{Kim2012}.  Thus, considering two magnon branches, we get \cite{Hess2003}:
\begin{equation}
\kappa_\mathrm{mag}=l_\mathrm{mag} \frac{2v_0k_B}{a^2c}\frac{T^2}{\Theta_M^2}\int \limits_{x_0}^{\Theta_M/T}x^2\sqrt{x^2-x_0^2}\frac{e^x\mathrm{d}x}{(e^x-1)^2},
\label{eq:1}
\end{equation}
with $x_0=\Delta/k_BT$ and $a=3.9$~\r{A}, $c=25.8$~\r{A} the relevant lattice parameters \cite{Crawford1994, Fujiyama2012}, and $\Theta_M=\hbar v_0 \sqrt{\pi}/ak_B$ the Debye temperature for magnons. 

We use Eq.~\ref{eq:1} to model the low temperature $\kappa_\mathrm{mag}$ by assuming a temperature independent $l_\mathrm{mag}$ for a certain temperature range. This corresponds to the physical picture of dominating magnon boundary scattering. A corresponding fit is shown in Fig.~\ref{fig:figure2} where we account for a possible offset $w$ due to an uncertain phonon background at low temperature \cite{Hess2003}. The fit describes the data well up to 75~K and yields a low temperature mean free path of $l_\mathrm{mag}=32\pm10$~nm. This value corresponds to $\sim82$ times of the Ir-Ir distance $a$.

%Using the known parameters of the low temperature fit, we calculate the temperature dependence of the mean free path $l_{mag}(T)$ for the whole measured temperature range using Eq.~(\ref{eq:1}). The resulting data are shown in Fig.~\ref{fig:figure3}. 
%At low temperature, $l_{mag}(T)$ is roughly constant up to $\sim75$~K reflecting the low-temperature boundary scattering as revealed by the afore analysis. For increasing temperature, $l_{mag}(T)$ decreases strongly where the change amounts up to an order of magnitude near room temperature.
%At the highest temperature accessible in this experiment, \fs{$l_\mathrm{mag}$ seems to saturate close to 4.6~nm which is slightly above the Ir-Ir distance, i.e., a natural minimum value of the mean free path}.

\begin{figure}
\includegraphics[clip,width=1\columnwidth]{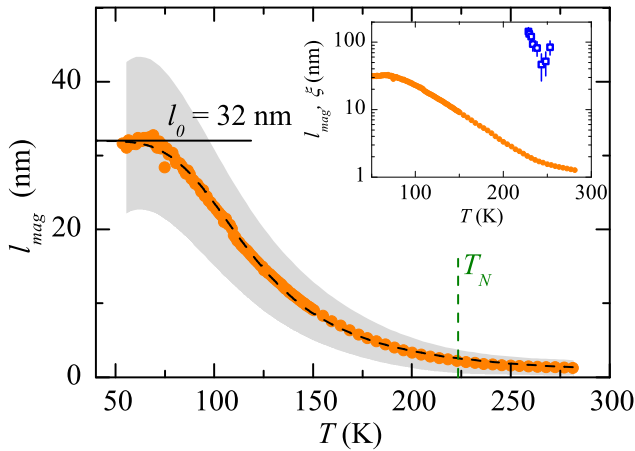}
\caption{Magnetic mean free path $l_\mathrm{mag}(T)$ of \sio (dots) resulting from the  phononic fit to $\kappa$ along $ab$. The shaded area marks the range spanned by the lower and upper bound for the phononic heat conductivity. At low temperatures the constant magnetic mean free path is found to be $l_0=(32\pm 10)$~nm. The dashed line is the phenomenological fit to $l_\mathrm{mag}(T)$ by Eq.~(\ref{matthiesen}) with $T^*=482$~K and $A_S=1.4\cdot10^7~\rm K^{-1}m^{-1}$.
The inset shows the magnetic mean free path together with the spin-spin correlation length $\xi$ (open squares) determined by RIXS \cite{Fujiyama2012} on a semilogarithmic scale.}
\label{fig:figure3}
\end{figure}

We now address the apparent deviation from a temperature-independent mean free path that becomes evident for higher temperatures $T\gtrsim75$~K. In this regime, the afore used simple low-temperature Debye approach cannot be employed anymore because it fails to properly describe the magnetic specific heat $c_\mathrm{mag}(T)$ at elevated temperatures. We therefore use the theoretical result of $c_\mathrm{mag}(T)$ of the $S=1/2$ Heisenberg antiferromagnet on a square lattice  \cite{Sengupta2003,Hofmann2003} with the exchange coupling $J=0.06$~eV \cite{Cetin2012,Hirata2013,Kim2012} and the above $v_0$ for calculating the temperature dependent $l_\mathrm{mag}(T)$ based on the kinetic expression~(\ref{eq:cvl}) for a 2D system. At low temperature (c.f. Fig.~\ref{fig:figure3}), $l_\mathrm{mag}(T)$ is roughly constant up to $\sim75$~K reflecting the low-temperature boundary scattering as revealed by the afore analysis. For increasing temperature, $l_\mathrm{mag}(T)$ decreases strongly where the change amounts up to an order of magnitude near room temperature. At the highest temperature accessible in this experiment, $l_\mathrm{mag}$ seems to saturate close to 1.2~nm, i.e., one order of magnitude above the Ir-Ir distance which constitutes a natural minimum value of $l_\mathrm{mag}$.

The strong decrease at elevated temperatures clearly signals the onset of a temperature-activated scattering process. We assume that both the low-temperature boundary scattering and the temperature-activated process are independent of each other. Following Matthiessen's rule, the mean free path can then be written as
\begin{equation}
l_\mathrm{mag}^{-1}=l_0^{-1}+\left(\frac{\exp(T^*/T)}{A_sT}\right)^{-1}, \label{matthiesen}
\end{equation}
where the second term is an empirical formula for temperature-activated scattering in magnetic heat transport that has been successfully used in one-dimensional systems \cite{Sologubenko2001,Hlubek2010,Hlubek2011,Hlubek2012} (with $T^*$ the characteristic energy scale of the temperature-activated scattering process and $A_s$ a proportionality factor). As can be seen in the figure, this formula fits the experimental $l_\mathrm{mag}(T)$ quite well. The fit yields $T^*\sim480$~K \footnote{We estimate the uncertainty of the fit result by separately fitting the upper and lower bounds of the grey shaded area, which yield $T^*=  450$~K with $A_S=0.8\cdot10^7~\rm K^{-1}m^{-1}$ and  $T^*=555$~K with $A_S=5.2*10^7\cdot10^7~\rm K^{-1}m^{-1}$.} which should be considered as a very coarse estimate of the energy of the most important scattering mode. The value roughly lies in the energy range of Ir-O-Ir bond bending  modes, which have been suggested to strongly couple to the electronic structure \cite{Moon2009,Cetin2012}. Thus, the primary cause of the temperature-activated scattering may be ascribed to the scattering of the magnons off these phonons.

It is instructive to compare the mean free path with the spin-spin correlation length $\xi$ measured by resonant x-ray diffusive scattering \cite{Fujiyama2012}, see inset of Fig.~\ref{fig:figure3}. This quantity is a conceivable natural upper limit for $l_\mathrm{mag}$. In the long-range ordered phase below \tn, the spin-spin correlation length is infinitely large and thus unimportant for the magnetic heat transport. For higher temperatures, $\xi(T)$ decreases rapidly with increasing temperature but remains still  more than an order of magnitude larger than $l_\mathrm{mag}$. Therefore, it plays only a minor role in limiting the magnetic heat conductivity above \tn, if any. This suggests that the seeming anomaly in $\kappa_{ab}$ around \tn (cf. Fig.~\ref{fig:figure1}) is mostly unrelated to the onset of magnetic ordering but rather connected with the growing importance of temperature-activated magnon scattering. Note, however, that a faint change of slope is discernible in the semilogarithmic representation of $l_\mathrm{mag}$ shown in the inset of Fig.~\ref{fig:figure3}.
% This finding is very similar to previous studies on 2D cuprate compounds, where also no anomaly in $\kappa_\mathrm{mag}$ is observed at \tn, and $l_\mathrm{mag}\ll\xi$ \cite{Hess2003,Hofmann2003}. 

% Thinking in terms of the strong spin-orbit coupling in the 5$d$ compounds the spins will most dominantly interact with the phonons. Thus, the temperature activated scattering can most probably be ascribed to phonon-magnon-coupling. This picture yields the following interpretation, that despite the strong spin-orbit coupled $J_\mathrm{eff}=1/2$ moments a significant magnetic heat conductivity is measured in \sio at low temperatures, because only boundary scattering limits the mean free path. At elevated temperatures phonons dominantly scatter the heat transporting magnons and suppress their mean free path strongly.

\begin{figure}
\includegraphics[clip,width=1\columnwidth]{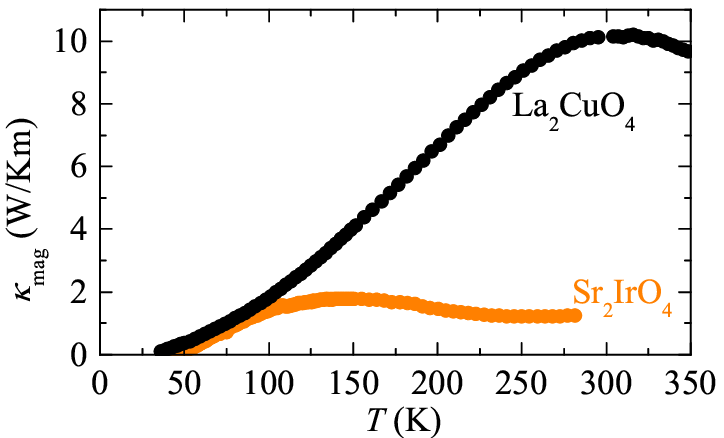}
\caption{Magnetic heat conductivity $\kappa_\mathrm{mag}$ of \sio in comparison with that of \lac (reproduced from Ref.~\onlinecite{Hess2003}).}
\label{fig:figure4}
\end{figure}

The strong magnon-phonon scattering evident in our data reveals a qualitative difference of the pseudospin heat transport of \sio and  the spin heat transport of the almost isostructural and thus closely related $S=1/2$-system \lac \cite{Hess2003}, as is inferred from a direct comparison, cf. Fig.~\ref{fig:figure4}. Both the spin wave velocity of \lac \cite{Hayden1991}, and the low-temperature magnon mean free path ($\sim 56$~nm) of the sample considered in the figure \cite{Hess2003} are approximately twice larger than those of our \sio sample. Considering these parameters and Eq.~(\ref{eq:1}), the almost identical low-temperature increase of both curves at $T\lesssim100$~K can be understood as the consequence of dominating magnon-boundary scattering in both cases. However, upon further increasing $T$, a strong suppression of $\kappa_\mathrm{mag}$ of \sio occurs while that of \lac continues to increase up to room temperature. Apparently, the magnon-phonon scattering in \sio is dramatically stronger than that in \lac, despite similar phonon spectra in both compounds \cite{Moon2009,Cetin2012,Pintschovius1989}. This unambiguously evidences a peculiar and particularly strong nature of the magneto-elastic coupling in \sio, arising from the large SOC and the resulting entanglement of spin and orbital degrees of freedom \cite{Jackeli2009}.

In conclusion, our data provide the first experimental result for low-dimensional $J_\mathrm{eff}=1/2$ pseudospin heat transport in an iridate compound. Our data show that the magnetic heat conductivity remains a valuable tool to probe the generation and the scattering of magnetic excitations also in these systems. At low temperatures $T\lesssim 100$~K, the magnetic heat transport $\kappa_\mathrm{mag}$ is  dominated by magnon scattering off static boundaries and thus comparable with that of 2D $S=1/2$ systems. However, at higher temperatures unusual strong magnon-phonon scattering becomes increasingly important, highlighting the peculiar nature of the pseudospin moments and excitations. 

% It thus is a promising approach for studying exotic elementary excitations in frustrated magnets like honeycomb or hyperkagome $J_\mathrm{eff}=1/2$ lattices.

\begin{acknowledgments}
This work was supported by the Deutsche Forschungsgemeinschaft through SFB 1143. Discussions with V. Kataev and W. Brenig are gratefully acknowledged.
\end{acknowledgments}

\end{document}